\documentclass[sigconf,edbt]{acmart-edbt2021}

\usepackage{amssymb} 
\usepackage{graphicx} 
\usepackage{enumitem} 
\usepackage{array} 
\usepackage{multirow}
\usepackage{amsfonts} 
\usepackage{subfigure} 
\usepackage{balance} 
\usepackage{booktabs} 

\long\def\comment#1{}

\setcopyright{rightsretained}
\acmDOI{}
\acmISBN{978-3-89318-084-4}
\acmConference[EDBT 2021]{24th International Conference on Extending Database Technology (EDBT)}{March 23-26, 2021}{Nicosia, Cyprus} 
\acmYear{2021}
\begin{document}

\title{SceneRec: Scene-Based Graph Neural Networks\\ for Recommender Systems}

\author{Gang Wang}
\affiliation{%
  \institution{SKLSDE Lab, Beihang University}
}
\email{iegwang@buaa.edu.cn}

\author{Ziyi Guo}
\affiliation{%
  \institution{JD.com}
}
\email{guoziyi@jd.com}

\author{Xiang Li}
\affiliation{%
  \institution{East China Normal University}
}
\email{lixiang3776@gmail.com}

\author{Dawei Yin}
\affiliation{%
  \institution{Baidu.com}
}
\email{yindawei@acm.org}

\author{Shuai Ma}
\affiliation{%
  \institution{SKLSDE Lab, Beihang University}
}
\email{mashuai@buaa.edu.cn}

\renewcommand{\shortauthors}{}

\begin{abstract}
Collaborative filtering has been largely used to advance modern recommender systems to predict user preference. 
A key component in collaborative filtering is representation learning, which aims to project users and items into a low dimensional space to capture collaborative signals. 
However, the scene information, which has effectively guided many recommendation tasks, is rarely considered in existing collaborative filtering methods. 
To bridge this gap, we focus on scene-based collaborative recommendation and propose a novel representation model SceneRec. 
SceneRec formally defines a \textit{scene} as a set of pre-defined item categories that occur simultaneously in real-life situations and creatively designs an item-category-scene hierarchical structure to build a scene-based graph. 
In the scene-based graph, we adopt graph neural networks to learn scene-specific representation on each item node, which is further aggregated with latent representation learned from collaborative interactions to make recommendations. 
We perform extensive experiments on real-world E-commerce datasets and the results demonstrate the effectiveness of the proposed method.


\end{abstract}

\maketitle              

\section{Introduction}
\label{sec:introduction}
Recommender systems have become increasingly important to address the information overload problem and have been widely applied in many different fields, such as social networks~\cite{yang2014survey} and news websites~\cite{ZhangC019}.  
To predict a user's preference, an extensive amount of collaborative filtering (CF) methods have been proposed to advance recommender systems. 
The basic idea of CF is that user behavior would always be similar and a user's interest can be predicted from the historical interactive data like clicks or purchases. 
A key component of CF is to learn the latent representation, which usually projects users and items into a lower dimensional space. 
A variety of CF models, including matrix factorization \cite{hu2008collaborative}, deep neural networks \cite{NCF_HeLZNHC17} and graph convolutional networks \cite{berg2017graph}, are adopted to capture collaborative signals from a user-item matrix or a user-item bipartite graph.

In the meantime, recommender systems that integrate scene information are attracting more and more attention. 
For example, predictive models are able to recommend substitutable or complementary items \cite{DBLP:conf/eccv/KiapourYBB14,DBLP:conf/cvpr/LiuLQWT16,kang2019complete} that visually match the scene which is represented in an input image. 
The image data contains rich contextual information like background color, location, landscape, etc., which may be ignored by conventional CF methods. 
However, the input image could reveal no scene information or even becomes unavailable in many recommendation scenarios. 
For example, in E-commerce systems, most thumbnail images only contain product pictures which are embedded in the white background. 
In such circumstances, scene-based recommendation becomes infeasible because the scene definition is not clear.

To address this issue, this work investigates the utility of incorporating scene information into CF recommendation. 
However, this study brings two challenges. 
First, a formal definition on scene is essential to this problem. 
Without image data, how to formally define a scene becomes a problem. 
Second, how to incorporate scene information into existing CF models should also be taken into account. 
Keeping these two key points in mind, we propose SceneRec, a novel method for scene-based collaborative filtering.
Specifically, we propose a principled item-category-scene hierarchical structure to construct the scene-based graph (Figure \ref{fig:network}). 
In particular, a \textit{scene} is formally defined by a set of fine-grained item categories that could simultaneously occur in real-life situations. 
For example, the set of item categories \{Keyboard, Mouse, Mouse Pad, Battery Charger, Headset\} represents the scene ``Peripheral Devices''. 
This can be naturally applied to a situation where a user has already bought a PC and many different types of supplementary devices are recommended. 
Moreover, SceneRec applies graph neural networks on the scene-based graph to learn the item representation based on the scene information, which is further aggregated with the latent representation learned from user-item interactions to make predictions.

To the best of our knowledge, SceneRec is among the first to study scene-based recommendation with a principled scene definition and our main contributions are summarized as follows:

\noindent
(1) We study the problem of scene-based collaborative filtering for recommender system where a scene is formally defined as a set of item categories that could reflect a real-world situation.


\noindent
(2)We propose a novel recommendation model SceneRec. 
It leverages graph neural networks to propagate scene information and learn the scene-specific representation for each item. 
This representation is further incorporated with a latent representation from user-item collaborative interactions to make predictions.

\noindent
(3) We conduct extensive experiments to evaluate the performance of SceneRec against 9 other baseline methods. 
We find that our method SceneRec is effective.
Specifically, SceneRec on average improves the two metrics (NDCG@10, HR@10) over the baselines by ($14.8\%, 12.1\%$) on 4 real-world datasets. 
\section{Related Work}

\label{sec:related}
Collaborative filtering has been widely applied in modern recommender systems. 
One class of CF methods try to build explicit models on the user-item interactions. 
For example, matrix factorization
\cite{hu2008collaborative,BPR_RendleFGS09,LaishramSPU16MfRec,Diaz-AvilesGN12MfRec}
maps the representation of each user and each item into a lower dimensional space and calculates inner product between vector representations to make predictions. 
To enhance recommendation, various contextual information has been incorporated into CF, 
such as user review \cite{xu2014collaborative}, social connections \cite{yang2014survey} and item side information \cite{wang2015collaborative}. 
Different from existing works that rely on linear predictive function, many recent efforts apply deep learning techniques
\cite{NCF_HeLZNHC17}
to learn non-linearities between user embedding and item embedding.

Another line of CF methods take user-item interactions as a bipartite graph. 
For example, some early efforts \cite{gori2007itemrank} conduct label propagation, which essentially searches neighborhood on the graph, to capture collaborative signals.
Inspired by the success of graph neural networks (GNN)
\cite{kipf2016semi,hamilton2017inductive} 
that directly conduct convolutional operations on the non-grid network data, a series of GNN-based recommendation methods have been proposed on an item-item graph \cite{PinSAGE_YingHCEHL18} or a user-item graph 
\cite{berg2017graph}
to learn a vector embedding for each item or user. 
The general idea is the representation of one graph node can be aggregated and combined by the representation of its neighbor nodes. 
NGCF \cite{NGCF_Wang0WFC19} extends GNN to multiple depths to capture high-order connectivities that are included in user-item interactions. 
KGAT~\cite{KGAT_Wang00LC19} and KGCN~\cite{KGCN_WangZXLG19} investigate the utility of incorporating knowledge graph (KG) into CF by projecting KG entities to item nodes.

Our work is also related to the application of scene information in recommender systems.
For example, given the scene in the form of an input image, recommendation methods are capable of providing substitutable \cite{DBLP:conf/eccv/KiapourYBB14,DBLP:conf/cvpr/LiuLQWT16} or supplementary\cite{kang2019complete} products that visually match the input scene.
However, in these tasks, the scene is represented by image data, which is not readily available in many recommendation scenarios. 
In such cases, scene-based recommendations become difficult or even impossible because the scene has not been well defined. 
In this paper, we aim to integrate scene information into CF where each scene is define by a set of fine-grained item categories. 
By exploiting the scene-specific representation into conventional CF signals, the model can potentially improve predictions on user preference.

\section{Problem Formulation}
\label{sec:preliminaries}


\begin{definition}
\textbf{Scene}. 
A \emph{scene} is defined as a set of item categories that occur simultaneously and frequently in a real-life situation, 
denoted as $s = \{ c_1, c_2, \cdots , c_{|s|} | c_i \in \mathcal{C}, 1 \leq i \leq |s| \}$, 
where $\mathcal{C}$ is the set of item categories and $|s|\geq 1$.
The item category is one of a item's attributes and $s \subset \mathcal{C}$.

\end{definition}


\begin{definition}

\textbf{User-Item Bipartite Graph.} 
The user-item interactions can be represented as a bipartite graph $\mathcal{G} = \{ (u, x_{ui}, i) | \\ u \in \mathcal{U}, i\in \mathcal{I}  \}$, 
where $\mathcal{U}$ and $\mathcal{I}$ are the set of users and items respectively, 
and the edge $x_{ui}$ indicates the occurrence or frequency with that the user $u$ has interacted with the item $i$,
such as clicking and purchasing.
\end{definition}

\begin{figure}[t]
    \centering
    \vspace{-2mm}
    \includegraphics[width=0.7\columnwidth]{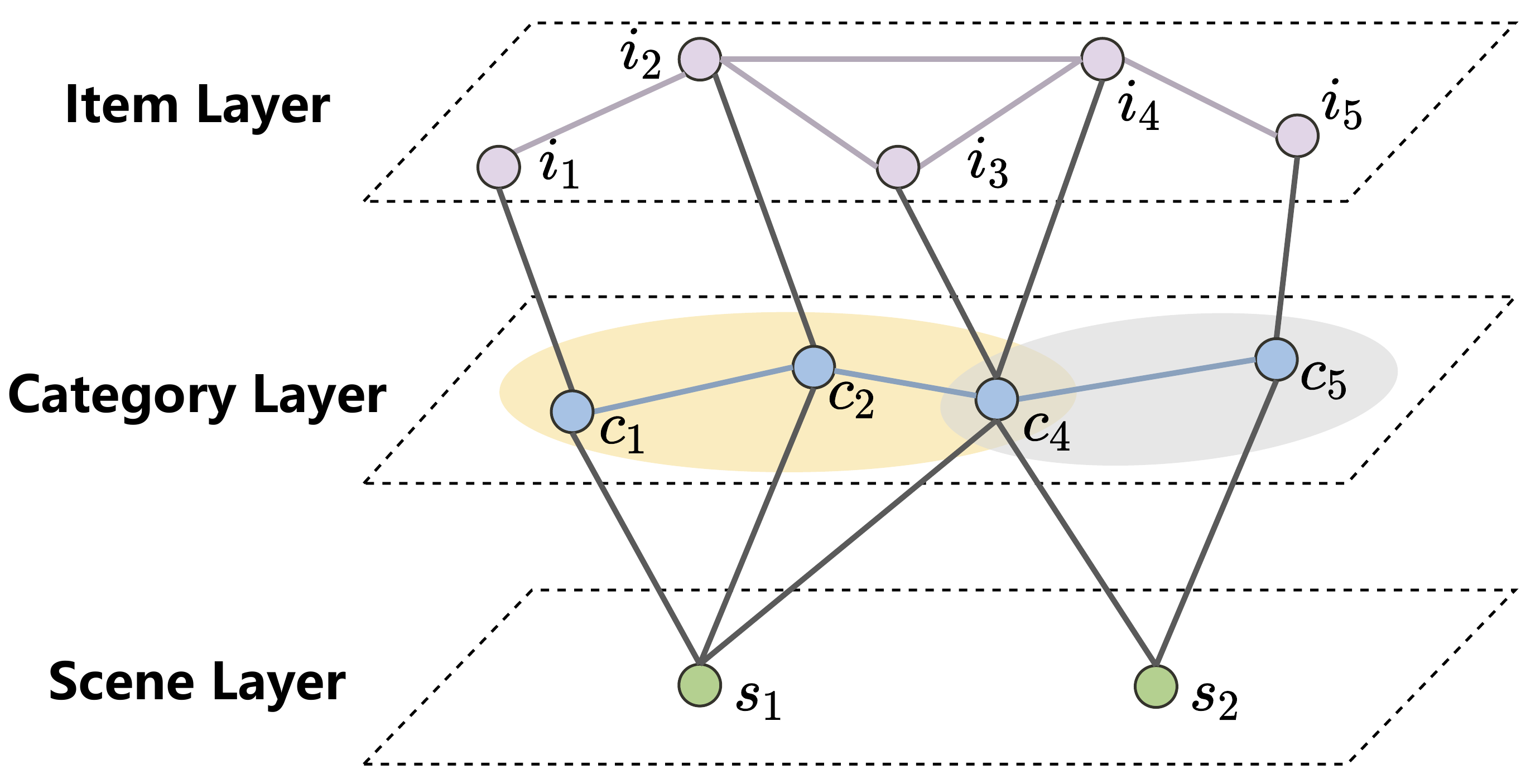}
    \vspace{-3mm} 
    \caption{An illustrative example of the scene-based graph that consists of the item layer, the category layer and the scene layer. 
    Each item is associated with a category. 
    In the item layer and the category layer, the set of edges represent the item-item relations and the category-category relations. 
    There are connections between categories and scenes, which indicates that a category belongs to a scene. }
    \label{fig:network}
    \vspace{-3mm}
\end{figure}

\begin{definition}
\textbf{Scene-based Graph.} 
The scene-based graph $\mathcal{H}$ is a hierarchical network with three layers: the item layer, the category layer, and the scene layer as shown in Figure~\ref{fig:network}.
The item layer consists of items and is denoted as $\mathcal{L}_{item} = \{ (i_p, y_{pq}, i_q) | i_p, i_q \in \mathcal{I}  \}$, 
where the edge $y_{pq}$ represents the similarity between two items $i_p$ and $i_q$. 
The category layer is denoted as $\mathcal{L}_{cate} = \{ (c_p, z_{pq}, c_q) | c_p, c_q \in \mathcal{C} \}$,
where the edge $z_{pq}$ represents that the category $c_p$ has relevance to the category $c_q$.
The interaction between the item layer and the category layer is described by $\mathcal{L}_{ic} = \{ (i_p, a_{pq}, c_q) | i_p \in \mathcal{I}, c_q \in \mathcal{C} \}$, 
where the edge $a_{pq}$ connects an item $i_p$ to a pre-defined item category $c_q$.
The scene layer is composed of scenes,
where a scene $s$ is formally defined as a set of item categories $\{c_1,c_2,\cdots, c_{\vert s \vert}\}$.
The relation between categories and scenes is illustrated by $\mathcal{L}_{cs} = \{ (c_p, b_{pq}, s_q) | c_p \in \mathcal{C}, s_q \in \mathcal{S} \}$, 
where the edge $b_{pq}$ indicates that a category $c_p$ belongs to a scene $s_q$ and $\mathcal{S} = \{s_1, s_2, \cdots \}$ is the set of scenes.
For simplicity,
we set the weights of edges in the scene-based graph $\mathcal{H}$ to be 1; otherwise, 0.
\end{definition}


\begin{definition}

\textbf{Scene-based Recommendation.} 
Given a user-item bipartite graph $\mathcal{G}$ recording interaction history, 
the goal of the scene-based recommendation is to predict the probability $\mathbf{r}_{ui}$ that the user $u$ has potential interest in the item $i$ with the help of scene information from a scene-based graph $\mathcal{H}$.
\end{definition}

\begin{figure*}[t]
    \vspace{-4mm} 
    \centering
    \includegraphics[width=\textwidth]{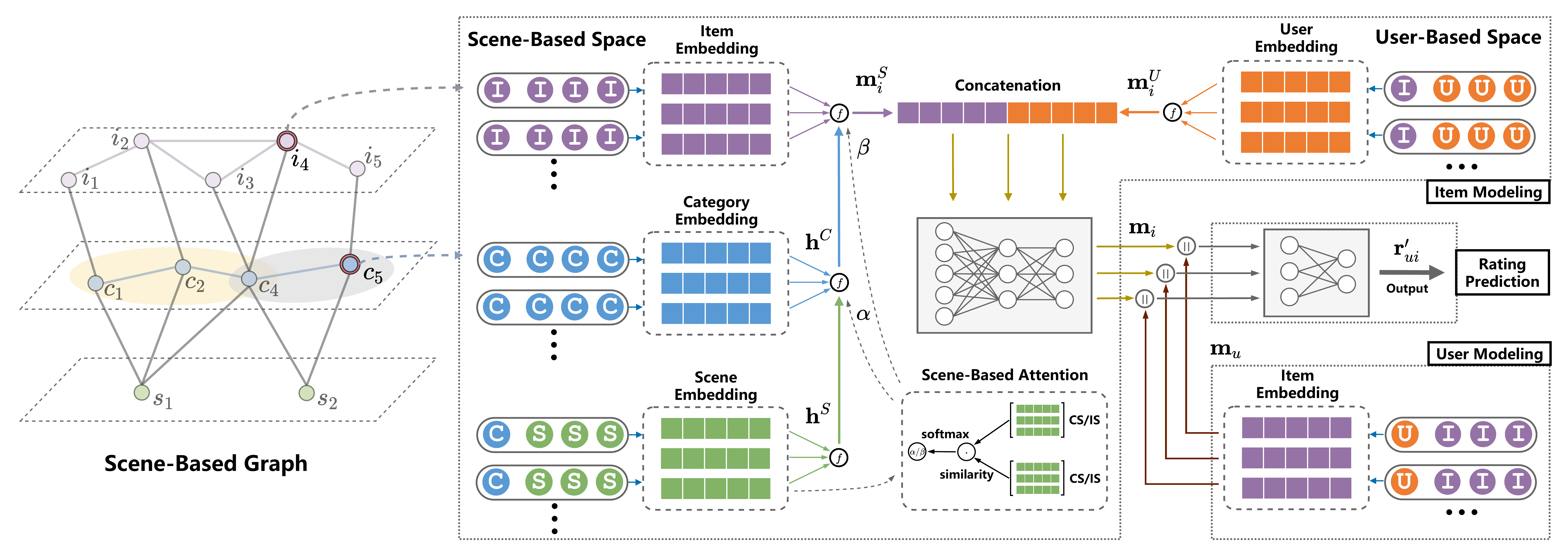}
    \vspace{-4mm}
    \caption{The illustration of SceneRec architecture (the arrowed lines present the bottom-up information flow). The embeddings of users and items are learned by user modeling and item modeling, respectively.}
    \label{fig:framework}
    \vspace{-2mm}
\end{figure*}

\section{Framework}
\label{sec:methods}
In this section, we will first give an overview about the proposed framework, then introduce each model component in detail.

\subsection {Architecture Overview}

The architecture of the proposed model is shown in Figure \ref{fig:framework}. 
There are three components in the model: user modeling, item modeling, and rating prediction. 
User modeling aims to learn a latent representation for each user. 
To achieve this, we take user-item interaction as input and aggregate the latent representation of items that the user has interacted with to generate the user latent factor. 
Item modeling aims to generate the item latent factor representation. 
Since each item exists in both user-item bipartite graph and the scene-based graph, SceneRec learns item representations in each graph space, i.e., item modeling in the user-based space and item modeling in the scene-based space. 
In the user-based space, we take a similar strategy which aggregates the representation of all users that each item has interacted with to generate vector embedding. 
In the scene-based space, we exploit the hierarchical structure of the scene-based graph where the information is propagated from the scene layer to the category layer and from the category layer to the item layer. 
Then we concatenate two item latent factors for the general representation. 
In the last component, we integrate item and user representations to make rating prediction.

\subsection {User Modeling}
In the user-item graph, a user $u_p$ is connected with a set of items and these items directly capture the user's interests.
We thus learn user $u_p$'s embedding $\mathbf{m}_{u_p}$ by aggregating the embeddings of item neighbors, which is formulated as, 
\begin{equation}
    \mathbf{m}_{u_p} = \sigma (\mathbf{W_u} \cdot \left\{ \sum_{i_q \in UI(u_p)} \mathbf{e}_{i_q} \right\} + \mathbf{b_u} ),
\end{equation}
where $UI(u_p)$ denotes the set of items that are connected to user $u_p$, $\mathbf{e}_{i_q}$ is the embedding vector of item $i_q$, and $\sigma$ is the nonlinear activation function. 
$\mathbf{W_u}$ and $\mathbf{b_u}$ are the weight matrix and the bias vector to be learned.

\subsection{Item Modeling}
The general representation $\mathbf{m}_{i_p}$ for item $i_p$ can be further split into two parts: the embedding $\mathbf{m}_{i_p}^U$ in the user-based space and the embedding $\mathbf{m}_{i_p}^S$ in the scene-based space. 

\subsubsection{User-based embedding} 
In the user-item graph, an item $i_p$ has connections with a set of users.
We learn its embedding $\mathbf{m}_{i_p}^U$ by aggregating the embedding of these engaged users:
\begin{equation}
    \mathbf{m}_{i_p}^U = \sigma (\mathbf{W_{iu}} \cdot \left\{ \sum_{u_q \in IU(i_p)} \mathbf{e}_{u_q} \right\} + \mathbf{b_{iu}} ),
\end{equation}
where $IU(i_p)$ denotes the set of users that are connected to item $i_p$, $\mathbf{e}_{u_q}$ is the embedding vector of user $u_q$,
$\mathbf{W_{iu}}$ and $\mathbf{b_{iu}}$ are parameters to be learned.
Since $\mathbf{m}_{i_p}^U$ is aggregated from user neighbors,
$\mathbf{m}_{i_p}^U$ represents the user-based embedding of item $i_p$.

\subsubsection{Scene-based embedding} 
In the scene-based graph, each item is connected to both other items and its category.
So, the scene-based embedding $\mathbf{m}_{i_p}^S$ for item $i_p$ is composed of representation that is specific to item neighbors and category neighbors.

For the category-specific representation, we should first generate the latent factor of each category. 
Since one category node can connect to both scene nodes and other related category nodes, the category representation can be further split into two types: the scene-specific and category-specific representation.

Given a category $c_p$, it may belong to a set of scenes and its scene-specific embedding vector $\mathbf{h}_{c_p}^S$ can be updated as follows:
\begin{equation}
    \mathbf{h}_{c_p}^S =  \sum_{s_q \in CS(c_p)} \mathbf{e}_{s_q},  
\end{equation}
where $CS(c_p)\!$ is the set of scenes that category $c_p$ belongs to and $\mathbf{e}_{s_q}\!$ is the embedding vector of scene $s_q$.

Besides the connection between scene nodes and category nodes, our model also captures the interactions between different category nodes. 
Each category contributes to the category-specific representation but categories do not always affect each other equally. 
Therefore, we apply the attention mechanism to learn the influence between different item categories. 
In this way, the category-specific representation $\mathbf{h}_{c_p}^C$ of the category $c_p$ can be aggregated as follows:
\begin{equation}
    \mathbf{h}_{c_p}^C =  \sum_{c_q \in CC(c_p)} \alpha_{pq} \mathbf{e}_{c_q}, 
\end{equation}
where $CC(c_p)$ is the set of neighbor categories, $\mathbf{e}_{c_q}$ is the embedding vector of $c_q$, and $\alpha_{pq}$ is the attention weight.
For a pair of categories, the more scenes they share, the higher relevance between them.
Therefore, we propose a scene-based attention function to compute $\alpha_{pq}$. 
Specifically, we calculate the attention score by comparing the sets of scenes that $c_p$ and $c_q$ belong to:
\begin{equation}
\label{eq:att_cat}
    \alpha_{pq}^{*} = f \left(  \sum_{s_a \in CS(c_p)} \mathbf{e}_{s_a}, \sum_{s_b \in CS(c_q)} \mathbf{e}_{s_b} \right), 
\end{equation}
where $f(\cdot)$ is an attention function to measure the input similarity. 
For simplicity, we use cosine similarity as $f(\cdot)$ in this work. 
$\alpha_{pq}$ is obtained by further normalizing $\alpha_{pq}^{*}$ via the softmax function:
\begin{equation}
\label{eq:att_cat2}
    \alpha_{pq}=\frac{\exp \left(\alpha_{pq}^{*}\right)}{\sum_{ \{q | \forall c_q \in CC(c_p)\} } \exp \left(\alpha_{pq}^{*}\right)}.
\end{equation}

Finally, we generate the overall representation $\mathbf{m}_{c_p}$ of category $c_p$ by integrating the scene-specific representation and the category-specific representation:
\begin{equation}
    \mathbf{m}_{c_p} = \sigma\left(\mathbf{W_{ic}} \cdot [\mathbf{h}_{c_p}^s \| \mathbf{h}_{c_p}^c] +\mathbf{b_{ic}} \right),
\end{equation}
where $\|$ denotes the concatenation operation, $\mathbf{W_{ic}}$ and $\mathbf{b_{ic}}$ are parameters to be learned.

For item $i_p$, it is only connected to one pre-defined category and thus its category-specific representation $\mathbf{h}_{i_p}^C$ is denoted as:
\begin{equation}
    \mathbf{h}_{i_p}^C =  \mathbf{m}_{C(i_p)},   
\end{equation}
where ${C(i_p)}$ indicates the category of $i_p$.

\begin{table*}[t]
    \centering
    \small
    \vspace{-4mm}
    \caption{Statistics of JD datasets. Each relation A-B has three parts: number of A, number of B, and number of A-B.} 
    \vspace{-2ex}
    \scalebox{0.9}{
        \begin{tabular}{ c | c | c | c | c}
        \toprule
             Relations (A-B) & Baby \& Toy & Electronics & Fashion & Food \& Drink  \\
        \midrule
            User-Item & 4,521-51,759 (481,831) & 3,842-52,025 (539,066) & 3,959-53,005 (541,238) & 3,236-47,402 (463,391)  \\
            Item-Item & 51,759-51,759 (3,002,806) & 52,025-52,025 (2,992,333) & 53,005-53,005 (2,750,495) & 47,402-47,402 (2,606,003)  \\
            Item-Category & 51,759-103 (51,759) & 52,025-78 (52,025) & 53,005-91 (53,005) & 47,402-105 (47,402)   \\
            Category-Category & 103-103 (1,791) & 78-78 (825) & 91-91 (1,058) & 105-105 (1,628) \\
            Scene-Category & 323-103 (1,370) & 54-78 (281) & 438-91 (1,646) & 136-105 (630) \\
        \bottomrule
        \end{tabular}
    }
    \label{tab:datasets}
    \vspace{-3ex}
\end{table*}

We continue to learn the item-specific representation $\mathbf{h}_{i_p}^I$
since there exist connections between different item nodes. 
Similar to category-category relations, items do not always affect each other equally and we apply the attention mechanism to learn $\!\mathbf{h}_{i_p}^I$:
\begin{equation}
    \mathbf{h}_{i_p}^I =  \sum_{i_q \in II(i_p)} \beta_{pq} \mathbf{e}_{i_q},   
\end{equation}
where $\beta_{pq}$ denotes the attention weight.
Since items that belong to the same category share similarity, we leverage scene information to calculate $\beta_{pq}$ by comparing their categories via the scene-based attention mechanism:
\vspace{-1mm}
\begin{equation}
\label{eq:att_item}
    \beta_{pq}^{*} = f \left(  \sum_{s_a \in IS(i_p)} \mathbf{e}_{s_a}, \sum_{s_b \in IS(i_q)} \mathbf{e}_{s_b} \right),
\end{equation}
\vspace{-2mm}
\begin{equation}
\label{eq:att_item2}
    \beta_{pq}=\frac{\exp \left(\beta_{pq}^{*}\right)}{\sum_{ \{q | \forall i_q \in II(i_p)\} } \exp \left(\beta_{pq}^{*}\right)},
\end{equation}
where $IS(i_p)$ is the set of scenes that contain item $i_p$'s category.

In the end, we concatenate the category-specific representation $\mathbf{h}_{i_p}^C$ and the item-specific representation $\mathbf{h}_{i_p}^I$ to derive the overall representation $\mathbf{m}_{i_p}^S$ of the item $i_p$ in the scene-based space:
\begin{equation}
    \mathbf{m}_{i_p}^S = \sigma\left(\mathbf{W_{ii}} \cdot [\mathbf{h}_{i_p}^C \| \mathbf{h}_{i_p}^I] + \mathbf{b_{ii}} \right),
\end{equation}
where $\mathbf{W_{ii}}$ and $\mathbf{b_{ii}}$ are parameters to be learned.

\subsubsection{The general item embedding}
The item embedding $\mathbf{m}_{i_p}^U$ in the user-based space learns the collaborative signals from user-item interactions, 
while the item embedding $\mathbf{m}_{i_p}^S$ in the scene-based space provides additional information from the scene-based graph. 
These two types of representations could be complementary to each other, and they are combined by a multilayer perceptron (MLP) to generate the general item embedding as follows:
\begin{equation}
    \mathbf{m}_{i_p} = {\mathbf{F}}\left(\mathbf{W_{i}} \cdot [ \mathbf{m}_{i_p}^U \| \mathbf{m}_{i_p}^S ] + \mathbf{b_{i}} \right), 
\end{equation}
where $\mathbf{F}(\cdot)$ is a MLP network, $\mathbf{W_{i}}$ and $\mathbf{b_{i}}$ are parameters. 

\subsection{Model Optimization}
Given the representation of user $u_p$ and the general representation of item $i_q$, the user preference is obtained via a MLP network:
\begin{equation}
    \mathbf{r}_{pq}' = {\mathbf{F}}\left(\mathbf{W_r} \cdot [ \mathbf{m}_{u_p} \| \mathbf{m}_{i_q} ] + \mathbf{b_r} \right), 
\end{equation}
where $\mathbf{W_{r}}$ and $\mathbf{b_{r}}$ are parameters to be learned.

To optimize the model parameters, we apply the pairwise BPR loss~\cite{BPR_RendleFGS09}, which takes into account the relative order between observed and unobserved user-item interactions and assigns higher prediction scores to observed ones. 
The loss function is as follow: 
\vspace{-2mm}
\begin{equation}
    \Omega(\Theta)=\sum_{(p, x, y) \in \mathcal{O}}-\ln \sigma\left( \mathbf{r}_{px}' - \mathbf{r}_{py}'\right) +\lambda\|\Theta\|_{2}^{2},
\end{equation}
where $\mathcal{O}=\left\{(p, x, y) |(p, x) \in \mathcal{R}^{+},(p, y) \in \mathcal{R}^{-}\right\}$ denotes the pairwise training data, 
$\mathcal{R}^{+}$ and $\mathcal{R}^{-}$ are the observed and unobserved user-item interactions, respectively.
$\Theta$ denotes all trainable model parameters and $\lambda$ controls $\ell_2$ regularization to prevent overfitting.

To sum up, we have different entity types, i.e., user, item, category and scene, in the user-item bipartite graph and the scene-based graph. 
In the learning process, the user representation is learnt from interactions between users and items. 
The item latent factor is generated from two components: the representation in the user-based space and the representation in the scene-based space. 
Then the user embedding and the item embedding are integrated to make prediction via pairwise learning.

\section{Experiments}
\label{sec:experiments}
In this section, we evaluate SceneRec on 4 real-world E-commerce datasets and focus on the following research questions:


\noindent
\textbf{RQ1}: How does SceneRec perform compared with state-of-the-art recommendation methods?

\noindent
\textbf{RQ2}: How do different key components of SceneRec affect the model performance?

\noindent
\textbf{RQ3}: How does the scene information benefit recommendation?  


\vspace{-1mm}
\subsection{Datasets}

To the best of our knowledge, there are no public datasets that describe scene-based graph for recommender systems. 
To evaluate the effectiveness of SceneRec, we construct 4 datasets, namely, Baby \& Toy, Electronics, Fashion, and Food \& Drink, from JD.com, one of the largest B2C E-commerce platform in China. 
In each dataset, we build the user-item bipartite graph and the scene-based graph from online logs and commodity information. 
Statistics of the above datasets are shown in Table \ref{tab:datasets} and more details are discussed next.


We first build the user-item bipartite graph that by randomly sampling a set of users and items from online logs. 
A user is then connected to an item if she or he clicked the item.

Next we build the scene-based graph where three different nodes, i.e., item, category and scene, are taken as input. 
We first consider connections between different item nodes. 
In E-commerce systems, users perform various behaviors such as ``view'' and ``purchase'', which can be further used to construct item-item relations. 
In this work, we choose ``view'' to build the item-item connections. 
A view session is a sequence of items that are viewed by a user within a period of time and it is intuitive that two items should be highly relevant if they are frequently co-viewed. 
In the item layer, two items are linked if they are co-viewed by a user within the same session where the weight is the sum of co-occurrence frequency within $2$ months. 
For each item, we rank all the connected items by the edge weight and at most top $300$ connections are preserved. 
All time period and numbers of connection are empirically set based on the trade-off between the size of datasets and co-view relevance between items.

\begin{table*}
    \centering
    \small
    \vspace{-4mm}
    \caption{Comparisons with baselines and model variants.}
    \vspace{-2ex}
    \label{tab:baselines}
    \centering
    \scalebox{0.9}{
    \begin{tabular}{c | p{1.5cm}<{\centering} p{1.5cm}<{\centering}  | p{1.5cm}<{\centering} p{1.5cm}<{\centering} | p{1.5cm}<{\centering} p{1.5cm}<{\centering}  | p{1.5cm}<{\centering} p{1.5cm}<{\centering}}
    \toprule
                   & \multicolumn{2}{c}{Baby \& Toy}            & \multicolumn{2}{c}{Electronics}& \multicolumn{2}{c}{Fashion}            & \multicolumn{2}{c}{Food \& Drink}\\
                   & NDCG@10 & HR@10  & NDCG@10 & HR@10  & NDCG@10 & HR@10  & NDCG@10 & HR@10 \\
    \midrule
    BPR-MF         & 0.3117  & 0.5213 & 0.4005  & 0.6082 & 0.3142  & 0.5294 & 0.3663  & 0.5445\\
    NCF            & 0.2232  & 0.3800 & 0.3324  & 0.5364 & 0.1518  & 0.3090 & 0.3068  & 0.4628\\
    CMN            & 0.2136  & 0.3840 & 0.4447  & 0.6725 & 0.2616  & 0.4516 & 0.4028  & 0.5854\\
    PinSAGE        & 0.2124  & 0.4145 & 0.2954  & 0.5200 & 0.1770  & 0.3724 & 0.2791  & 0.4798\\
    NGCF           & 0.3679  & 0.6000 & 0.4308  & 0.6559 & 0.3361  & 0.5749 & 0.3487  & 0.5228\\
    KGAT           & 0.3055  & 0.5421 & 0.3616  & 0.6172 & 0.3115  & 0.5580 & 0.3221  & 0.5093\\
    \midrule
    SceneRec-noitem& 0.3977  & 0.6475 & 0.4748  & 0.7007 & 0.3936  & 0.6454 & 0.4080  & 0.6029\\
    SceneRec-nosce & 0.4193  & 0.6617 & 0.4715  & 0.7156 & 0.3933 & 0.6499 & 0.4156  & 0.6074\\
    SceneRec-noatt & 0.3950  & 0.6357 & 0.4665  & 0.7053 & 0.3953  & 0.6410 & 0.4138  & 0.6154\\
    \midrule
    SceneRec       & \textbf{0.4298} & \textbf{0.6771} & \textbf{0.4926} & \textbf{0.7524} & \textbf{0.4220} & \textbf{0.6763} & \textbf{0.4266} & \textbf{0.6211} \\
    \bottomrule
    
    \end{tabular}
    }
    \vspace{-3ex}
\end{table*}

We then connect each item to its pre-defined category to build the item-category relations.
We also consider connections between different category nodes as shown in the second layer of the scene-based graph. 
For example, in E-commerce systems, the category ``Mobile Phone'' is strongly related to the category ``Phone Case'' but has little relevance to the category ``Washing Machine'', and thus the first two categories are linked. 
To achieve this, we compute the co-view frequency within six months between each pair of category node, and only top $100$ connections of each category is preserved. 
In the end, each pair is further labeled as $0$ or $1$ from consensus decision-making by three data labeling engineers to indicate if there exists relevance or not.



%

The last step of building the scene-based graph is to link category nodes to scene nodes. 
Each scene consists of a set of selected categories which can be manually coded by human experts (scene mining is our future work). 
Specifically, this procedure consists of two steps. 
First, an expert team (about $10$ operations staff) edits a set of scene candidates based on the corresponding domain knowledge.
Then, a data labeling team which consists of $3$ engineers refines the generated scenes based on the criteria that whether each scene is reasonable to reflect a real-life situation. 

To sum up, there is a user-item bipartite graph and a scene-based graph in the constructed E-commerce datasets where we have different types of nodes, i.e., user, item, category and scene. 
The scene-based graph presents a 3-layer hierarchical structure. 
There exist multiple relations among items, categories and scenes that are derived from user behavior data, commodity information and manual labeling. 
Thus, the datasets have all the characteristics of networks we want to study as described in Section \ref{sec:preliminaries}. 

\subsection{Baselines}

SceneRec leverages scene information to learn the representation vector of users and items in recommendation. 
Therefore, we compare SceneRec against various recommendation methods or network representation learning methods.

\noindent
(1) \textbf{BPR-MF}~\cite{BPR_RendleFGS09}
is a benchmark matrix factorization (MF) model which takes the user-item graph as input and BPR loss is adopted.

\noindent
(2) \textbf{NCF}~\cite{NCF_HeLZNHC17}
leverages multi-layer perceptron to learn non-linearities between user and item interactions in the traditional MF model.

\noindent
(3) \textbf{CMN}~\cite{CMN_EbesuSF18}
is a state-of-the-art memory-based model to capture both global and local neighborhood structure of latent factors. 

\noindent
(4) \textbf{PinSAGE}~\cite{PinSAGE_YingHCEHL18}
learns node representations on the large-scale item-item network where the representation of one item can be aggregated by the representation of its neighbor nodes. 
Here, we directly apply PinSAGE on the input user-item bipartite graph.

\noindent
(5) \textbf{NGCF}~\cite{NGCF_Wang0WFC19}: 
This is a state-of-the-art GNN-based recommendation method, which learns the high-order connectivities based on the network structure.

\noindent
(6) \textbf{KGAT}~\cite{KGAT_Wang00LC19}
investigates the utility of KG into GNN-based collaborative filtering where each item is mapped to an entity in KG. 
In our experiments, we regard each scene as a special type of KG entity and link it to item nodes via the category node connection. 
In such cases, the scene-based graph is degraded to the one that contains only item-scene connections. 
The graph contains two types of relations: an item \textit{belongs to} a scene and a scene \textit{includes} an item.


\noindent
(7) \textbf{SceneRec-noitem} 
is a variant of SceneRec by removing item-item interactions in the scene-based graph.

\noindent
(8) \textbf{SceneRec-nosce}
is a variant of SceneRec by removing both category and scene nodes, and thus the scene-based graph only includes relations between items.

\noindent
(9) \textbf{SceneRec-noatt}
is another variant of SceneRec by removing the attention mechanism between item-item relations and category-category relations.

\subsection{Experimental Settings}

We evaluate the model performance using the leave-one-out strategy as in~\cite{chen2017attentive,NCF_HeLZNHC17}. 
For each user, we randomly hold out one positive item that the user has clicked and sample $100$ unobserved items to build the validation set. 
Similarly, we randomly choose another positive item along with $100$ negative samples to build the test set. 
The remaining positive items form the training set.

In our experiments, we choose \textit{Hit Ratio} (HR) and \textit{Normalized Discounted Cumulative Gain}~\cite{rsh2011} (NDCG) as evaluation metrics. 
HR measures whether positive items are ranked in the top $K$ scores while NDCG focuses more on hit positions by assigning higher scores to top results. 
For both metrics, a larger value indicates a better performance.
We report the average performance over all users with $K=10$.








The hyper-parameters of SceneRec are fine-tuned using the validation set. 
We apply RMSProp~\cite{book_DL} as the optimizer where the learning rate is determined by a grid search among $\{\!10^{-4}\!,\! 10^{-3}\!,\! 10^{-2}\!,\! 10^{-1}\!\}$ and the $\ell_2$ normalization coefficient $\lambda$ is determined by a grid search among $\{0, 10^{-6}, 10^{-4}, 10^{-2} \}$. 
For fair comparisons, the embedding dimension $d$ is set to $64$ for all methods except NCF. 
For NCF, $d$ is set to $8$ due to the poor performance in higher dimensional space. 
For NGCF and KGAT, the depth $L$ is set to $4$ since it shows competitive performance via the high-order connectivity.

\subsection{Experimental Results}
\subsubsection{Performance Comparison (RQ1).}
Table~\ref{tab:baselines} reports comparative results of SceneRec against all $6$ baseline methods, and we have the following observations:




\noindent
(1) In general, NGCF achieves better results than baseline methods that take the user-item bipartite graph as input. There are two main reasons. 
First, GNN can effectively capture the non-linearity relations from user-item collaborative behaviors via information propagation on the graph. 
Second, NGCF learns the high-order connectivities between different types of nodes as shown in \cite{NGCF_Wang0WFC19}.

\noindent
(2) KGAT further adds KG information into recommender systems, but it does not obtain the best result. 
Note that the KG quality is essential to the model performance. 
In our work, there are no available KG attributes that match our datasets, so there is no additional information to describe network items. 
Furthermore, the simple item-scene connection loses rich relations, e.g. category-category interactions and item-item interactions, in the scene-based graph, and may not advance model prediction.

\noindent
(3) The proposed framework SceneRec obtains best overall performance using different evaluation metrics.
Specifically, SceneRec boosts ($16.8\%$, $10.8\%$, $25.6\%$, $5.9\%$) for NDCG@10, and ($12.9\%$, $11.9\%$, $17.6\%$, $6.1\%$) for HR@10 on datasets (Baby \& Toy, Electronics, Fashion, and Food \& Drink), compared with the best baseline.
There are several main reasons. 
First, SceneRec considers multiple types of entity nodes. 
To be specific, SceneRec generates embedding representations of users and items from the user-item bipartite graph while it learns complementary representations of items from the scene-based graph, which is not accessible in baseline methods. 
Second, SceneRec creatively designs a principled hierarchical structure in the scene-based graph where additional scene-guided information is propagated into collaborative filtering. 
Third, SceneRec leverages GNN which captures local network structure to learn non-linear transformation of different types of graph nodes. 
Fourth, SceneRec adopts attention mechanism to attentively learn weighting importance among item-item connections and category-category connections.


     

\subsubsection{Key Component Analysis (RQ2).}
Table~\ref{tab:baselines} also reports comparative results against 3 variants and it is observed that:

\begin{figure}[t]
    \vspace{-2mm}
    \centering
    \includegraphics[width=1\columnwidth]{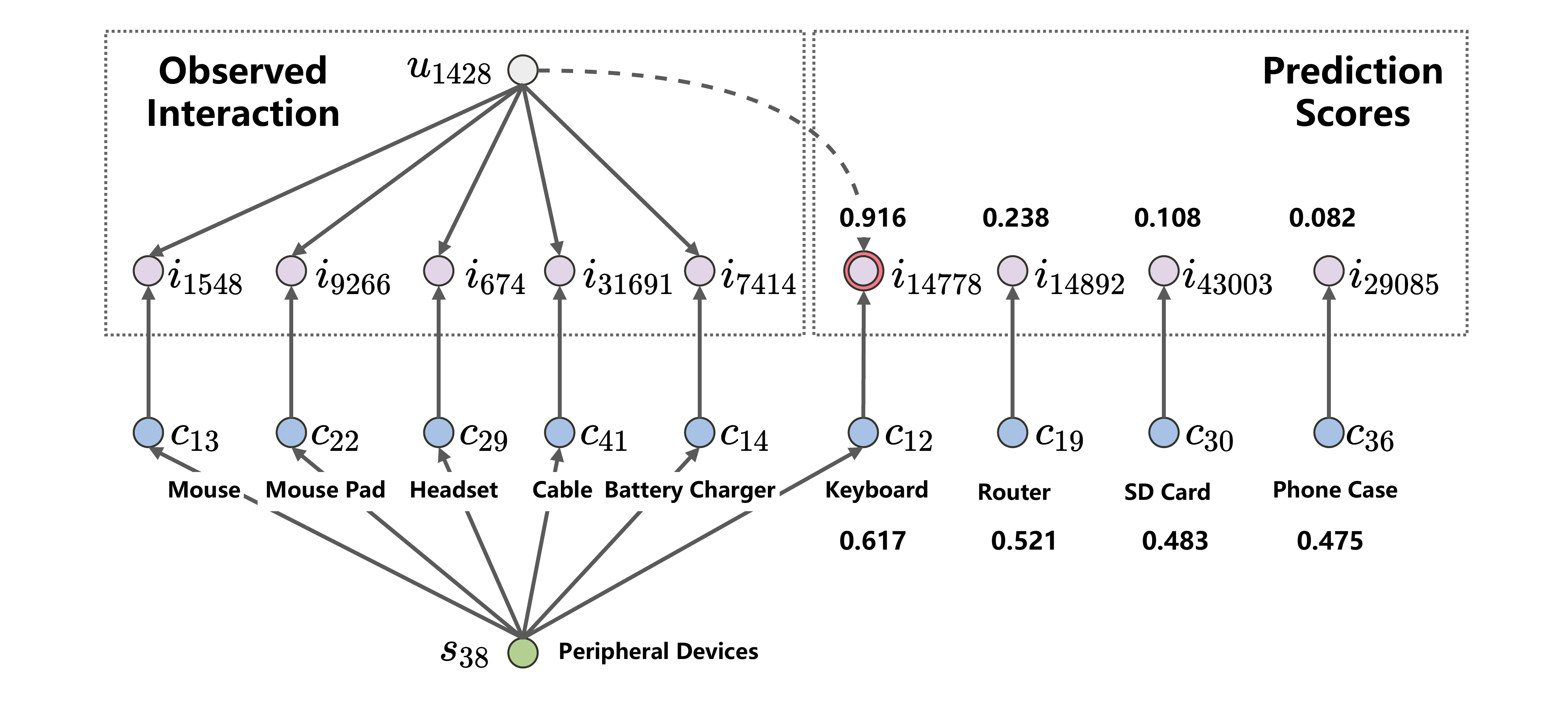}
    \vspace{-9mm}
    \caption{A real example on the Electronics dataset.}
    \label{fig:case}
    \vspace{-5mm}
\end{figure}


\noindent
(1) SceneRec-noitem obtains better experimental results than other baseline methods, and this indicates that the hierarchical structure of the scene-based graph can effectively propagate information and generate complementary scene-based representations. 
Moreover, SceneRec outperforms SceneRec-noitem and this verifies the effectiveness of incorporating item-item sub-network into the scene-based graph.

\noindent
(2) SceneRec-nosce outperforms all baselines because the item-item connections provide additional knowledge into conventional collaborative filtering. 
Comparing to SceneRec-nosce, SceneRec achieves better performance on both datasets and this indicates that, by leveraging scene information, SceneRec is capable of learning complementary representations beyond CF interactions.

\noindent
(3) The prediction result of SceneRec is consistently better than that of SceneRec-noatt, and this verifies that the attention mechanism does benefit the recommendation by learning weights of 1-hop neighbors for each item node or each category node.

\subsubsection{Case Study (RQ3).}
Finally, we use a case study to show the effects of integrating scene-specific representations into collaborative filtering in Figure~\ref{fig:case}. 
From the Electronics dataset, we randomly select a user $u_{1428}$, a set of items that the user has interacted with and a set of candidate items (whose prediction scores are given above item nodes). %
It is noted that we especially compute the average attention score (below the category node) between the candidate item and each item that the user has interacted with by the scene-based attentive mechanism.

The higher average attention score means more shared scenes between the candidate item and the user's interacted items. 
Therefore, the candidate item is more likely to occur in the scene derived from user interests, which could boost recommendation prediction.
From this case study, we see that the average attention score does relate to the prediction result. 
For example, the positive sample of item $i_{14778}$ that the user has interacted with has the highest prediction score and the largest average attention weight. 
Similar results can be also observed from other users. 
The item $i_{14778}$ is recommended because its category ``Keyboard'' complements the user-interacted items' categories in the same scene ``Peripheral Devices''.


\section{Conclusions}
\label{sec:conclusion}

In this paper, we investigate the utility of integrating the scene information into recommender systems using graph neural networks, where a scene is formally defined as a set of pre-defined item categories.
To integrate the scene information into graph neural networks, we design a principled 3-layer hierarchical structure to construct the scene-based graph and propose a novel method SceneRec. 
SceneRec learns item representation from the scene-based graph, which is further combined with the conventional latent representation learned from user-item interactions to make predictions. 
We conduct extensive experiments on four datasets that are collected from a real-world E-commerce platform. 
The comparative results and a case study demonstrate the rationality and effectiveness of SceneRec. 


\begin{acks}
This work is supported in part by National Key R\&D Program of China 2018AAA0102301 and NSFC 61925203.
\end{acks}


\bibliographystyle{ACM-Reference-Format}
\bibliography{paper_references}

\end{document}